\documentclass[12pt]{article}
\usepackage{graphics,epsfig}
\usepackage{amssymb}
\usepackage{amsmath}
\usepackage{epsfig}
\usepackage{url}

\newcommand \be {\begin{equation}}
\newcommand \ee {\end{equation}}
\newcommand \ba {\begin{array}}
\newcommand \ea {\end{array}}
\newcommand \bea{\begin{eqnarray}}
\newcommand \eea{\end{eqnarray}}
\newcommand \comphep{{\tt CompHEP}}
\newcommand \fpmc{{\tt FPMC}}
\newcommand \pomwig{{\tt POMWIG}}
\newcommand \herwig{{\tt HERWIG}}
\newcommand \herwigcc{{\tt HERWIG++}}
\newcommand \pt{\mbox{$p_T$} }

\newcommand{\units}[1]{\,\hbox{#1}}

\newcommand{\GeV}{\units{GeV}}

\newcommand{\pom}{\mathbb{P}}

\newcommand{\dkap}{\Delta\kappa^{\gamma}}

\newcommand{\lam}{\lambda^{\gamma}}
\newcommand{\aOw}{a_0^W}
\newcommand{\aOz}{a_0^Z}
\newcommand{\aCw}{a_C^W}
\newcommand{\aCz}{a_C^Z}
\newcommand{\wwgamma}{WW\gamma}
\newcommand{\cep}[1]{$gg \rightarrow #1$}
\renewcommand{\d}{\mathrm{d}}

\def\iproc{\url{IPROC}}
\def\nflux{\url{NFLUX}}
\def\typepr{\url{TYPEPR}}
\def\typint{\url{TYPINT}}
\newcommand{\refeq}[1]{(\ref{#1})}

\def\colfirst{4.5cm}

\newcommand{\hspaceID}{\hspace{5.2cm}}
 
\begin{document}

\begin{flushright}
\today
\end{flushright}

\vspace*{30mm}

\begin{center}
{\LARGE \bf FPMC : a generator for forward physics}

\par\vspace*{20mm}\par

{\large \bf M.~Boonekamp$^a$, A.~Dechambre$^a$, V.~Juranek$^b$, O.~Kepka$^b$, M.~Rangel$^c$, C.~Royon$^a$, R.~Staszewski$^d$}

\bigskip

{\em $^a$ 
CEA/IRFU/Service de physique des particules,\\ CEA/Saclay, 91191 Gif-sur-Yvette cedex, France}\\
{\em $^b$ 
Center for Particle Physics, Institute of Physics, Academy of Science, Prague }\\
{\em $^c$ 
Universidade Federal do Rio de Janeiro (UFRJ), Rio de Janeiro, Brazil}\\
{\em $^d$ Institute of Nuclear Physics, Polish Academy of Sciences, Krakow \\
}
\vspace*{5mm}

{\em E-mail:} maarten.boonekamp@cern.ch \\
              oldrich.kepka@cern.ch\\
              rafal.staszewski@ifj.edu.pl 

\vspace*{5mm}

\end{center}
\vspace*{15mm}

\begin{abstract}
We present the Forward Physics Monte Carlo (\fpmc) designed to simulated central particle production with
one or two leading intact protons and some hard scale in the event. 
The underlying interaction  between protons or anti-protons  through singlet exchange
can manifest itself in many forms. The following production mechanisms are implemented: single diffractive dissociation, 
double pomeron exchange, and exclusive production due to two-gluon or two-photon exchanges.
With increasing beam center-of-mass-energies, the production of new final states become possible
at the LHC.  The aim of \fpmc{} is to implement these processes in one common framework. 
\end{abstract}

\newpage

\section*{Program summary}

\noindent {\it Title of the program:} {\tt FPMC}, version {\tt 1.0}\\

\noindent {\it Computer:} any computer with the FORTRAN 77 or GFORTRAN compiler under the UNIX or Linux operating systems.\\

\noindent {\it Operating system:} UNIX; Linux\\

\noindent {\it Programming language used:} FORTRAN 77\\

\noindent {\it High speed storage required:} $<$ 100 MB \\

\noindent {\it Keywords:} Proton-(anti-)proton collisions, diffraction, exclusive production,
double pomeron exchange, two photon exchange.\\   

\noindent {\it Nature of the physical problem:} Proton diffraction at
               hadron colliders can manifest itself in many forms, and
               a variety of  
               models exist that attempt to describe it
               \cite{Ingelman:1984ns,Cahn:1990jk,Drees:1989vq,Papageorgiu:1995eg,Budnev:1974de,Khoze:2001xm,Cudell:2008gv}. This program
               implements some of the more  
               significant ones, enabling the simulation of central
               particle production through color singlet exchange
               between interacting  
               protons or anti-protons.\\

\noindent {\it Method of solution:} The Monte-Carlo method is used to
               simulate all elementary $2\rightarrow 2$ and
               $2\rightarrow 1$ processes available 
               in {\tt HERWIG}. The color singlet exchanges implemented
               in {\tt FPMC} are implemented as functions re-weighting
               the photon flux already  
               present in {\tt HERWIG}.\\

\noindent {\it Restriction on the complexity of the problem:} The
                program relying extensively on {\tt HERWIG}, the
                limitations are the same as in  
                \cite{Corcella:2002jc}.\\

\noindent {\it Typical running time:} Approximate times on a 2.5 GHz
               Dual-Core Intel: 1-60 minutes per 10000 unweighted events,
               depending on the 
               process under consideration.\\

\noindent {\it Homepage:} \url{www.cern.ch/fpmc}\\

\newpage

\section{Introduction}

In this paper, we present a Forward Physics Monte Carlo (FPMC) \cite{fpmc} generator to simulate inelastic processes occurring in hadron-hadron or
collisions in which one or
both hadrons stay intact. The focus of FPMC on the other hand is to simulate processes with a large mass produced in the central
pseudo-rapidity. This allows to apply perturbative methods to obtain predictions
for productions of electroweak boson, di-jets, Higgs boson, dilepton pairs etc.
On the contrary, the soft diffractive part of the cross section with in general low-\pt particle
production are implemented in other generators.

\par There are in general two types of processes with leading hadrons
distinguished in the diffractive community: \textit{exclusive} and
\textit{inclusive}. In exclusive events, empty regions in pseudo-rapidity called rapidity gaps
separate the intact very forward proton from the central massive object (e.g. di-jets). Exclusivity means that nothing else is produced except the leading protons and the central object. The
exclusive processes is due to underlying multi-gluon~\cite{Khoze:2001xm,Cudell:2008gv,Maciula:2010tv} or two-photon exchanges~\cite{Cahn:1990jk,Drees:1989vq,Papageorgiu:1995eg,Budnev:1974de} which we denote here as QCD and QED productions, respectively.  \par The inclusive processes
also exhibit rapidity gaps; however, in addition they contain soft particles
accompanying the production of a hard diffractive object and the rapidity gaps are subsequently
in general smaller than in the exclusive case. Measurements of these
processes have been successfully described by Ingelman-Schlein model~\cite{Ingelman:1984ns} which
involves exchanges of one or more perturbative pomerons.  The pomeron structure
is described by the parton distribution functions (PDF) measured in
events where rapidity gap or the intact leading proton is observed, mainly 
at HERA.  \par Hard
diffractive or exclusive physics has been studied in the past at various
colliders. Model predictions were obtained using many different generators
\cite{pompyt,Cox:2000jt,Boonekamp:2003ie,Monk:2005ji,Baranov:1991yq}. This is a first attempt to consolidate
predictions involving intact beam particles in the final state into one common
framework in order to ease MC prediction for upcoming forward physics program
at the LHC.  
\par The paper is organized as follows:
A program overview is given in Section~\ref{sec:overview}. The event in information
is discussed in Section~\ref{sec:eventinfo}. The main part of the paper is contained
in Section~\ref{sec:procover} where a description of processes that can be studied 
with \fpmc{} is given. The current work is concluded in Section~\ref{sec:conclusion}. 
In the Appendix details concerning the parameter setup to run \fpmc{} are provided.

\section{Program Overview}
\label{sec:overview}

The generation of forward processes is implemented inside \herwig{} version
6.500~\cite{Corcella:2002jc}. The original code simulating two-photon exchanges in
$e^+e^-$ collisions is adapted such that the pomeron/gluon is exchanged instead
of a photon and a particular proton structure in diffractive events is used in
hadron collision in this case. Note that such approach has first been applied in \pomwig~\cite{Cox:2000jt}.
 The user selects a particular model of interest  by
main steering parameter {\tt NFLUX }. The nature of the process is further
specified by parameters {\tt TYPEPR} to distinguish exclusive/inclusive
processes and  {\tt TYPINC} characterized the QED/QCD type of the exchange (see
Table~\ref{tab:nflux}). 
 \par
In certain processes which use standard \herwig{} non-diffractive matrix elements, 
the \herwig{} process numbering scheme is followed. In addition, new processes have
been added for example for the case of exclusive productions. For all processes 
simulated with \fpmc, the numbering should start with 10000. Adding 10000 
to the \herwig{} process code \iproc{} suppresses the underlying event production 
formation from beam remnants soft scattering. This is equivalent to setting 
{\tt PRSOF}=0. For more details see \herwig{} manual~\cite{Corcella:2002jc}.

In order to prevent interference with the
standard \herwig{} processes numbering, the \fpmc{} process numbers start from 10000.
The standard \herwig{} matrix elements are used in some cases. The corresponding 
process number in FPMC is the process number in \herwig{} plus 10000. 
In this cases, the details of the generation are steered by the 
same parameters that are used to control the production in \herwig.

\begin{table}[h]
  \begin{center}
  	\begin{tabular}{|r|l|} 
    	\hline
			\textbf{NFLUX} & \textbf{Description} \\ \hline  
			9  &  QCD factorized model, Pomeron flux \cite{Ingelman:1984ns}\\ 
			10 &  QCD factorized model, Reggeon flux \cite{Ingelman:1984ns}\\ 
			12 &	QED flux from Cahn, Jackson; $R\sim1.2A^\frac{1}{3}$~\cite{Cahn:1990jk} \\ 
			13 & 	QED flux from Drees et al., valid for heavy ions only~\cite{Drees:1989vq}\\ 
			14 & 	QED flux in pp collisions, from Papageorgiou~\cite{Papageorgiu:1995eg}\\ 
			15 & 	QED flux in pp collisions, from Budnev et al.~\cite{Budnev:1974de}\\ 
			16 & 	QCD KMR Exclusive model~\cite{Khoze:2001xm} \\
			17 & 	QCD CHIDe Exclusive model~\cite{Cudell:2008gv} \\ \hline
    	\hline
			\textbf{TYPEPR} & \\ \hline  \hline
			INC & Inclusive reaction  \\ 
			EXC	&	Exclusive reaction (only color-singlet amplitude) \\ \hline
    	\hline
			\textbf{TYPINT} &  \\ \hline  \hline
			QED&	Photon initiated process \\ 
			QCD &	Gluon/quark initiated process\\ \hline
		\end{tabular}
\end{center}
\caption{Main switches of the program that select implemented models of the forward
physics with leading protons.}
\label{tab:nflux}
\end{table}

\section{Event Information} 
\label{sec:eventinfo}

The \fpmc{} event information such as for example particle numbering, particle state, kinematics, 
particle production flow is  the same as in original \herwig{} and can be found
in  \herwig{} manual~\cite{Corcella:2002jc}. The following changes have been made. Because the implementation of 
the diffractive inclusive and exclusive processes in hadron-hadron collision is based on two-photon exchanges in
$e^+e^-$ collisions, the event record fixing is done in which the beam electrons are replaced 
by diffractive protons. Moreover, radiated photons off electrons are replaced by Pomerons/Reggeons where necessary 
(routine {\tt HWFXER}). New particle numbers are introduced for Pomeron and Reggeons, {\tt ID=990,110}, respectively.

\section{Process Overview}
\label{sec:procover}

The list of processes and their brief physics description is provided in the following sub-sections. The process numbers  
are given one. The user is invited to consult the full generator setup summary in Appendix~\ref{appa}.
\label{fixit: should also discuss xi cut, Q2 cut etc here as it is common }


\subsection{Central Exclusive QED Production - $\gamma\gamma$ Interactions}
\label{sec:excQED}

When an energy of relativistic  protons/leptons beams is sufficiently high,
 a hard process can be initiated by collinear photons emission. Subsequently, a large set of final states can be produced in photon-photon fusion (a photoproduction process with one
broken proton can also occur, but this process is not available in \fpmc).
\par
The two-photon interactions in $pp$\footnote{EPA describes production of photons for any charged particle for which the electromagnetic form factors are known.} collisions are described within the
Equivalent Photon Approximation (EPA) framework \cite{Budnev:1974de}. The cross section
is expressed in terms of the photon flux $f(\omega, q^2)$. It corresponds 
to the probability that a proton emits a photon of energy $\omega$ and 
momentum transfer $q^2$. Since the typical transverse momentum of the photon is very
small in two-photon interactions, $q^2$ dependence of the photon flux can be integrated out and the
 total cross section can be written as a product of the sub-process  $\gamma\gamma\rightarrow X$ cross section and the photon fluxes
\begin{equation}
      \frac{\d\sigma}{\d\Omega}=\int\d \omega_1\d \omega_2\frac{\d \sigma_{\gamma\gamma\rightarrow X}
(W=\sqrt{(4\omega_1\omega_2)})}{\d\Omega}f(\omega_1)f(\omega_2). 
\label{eq:totcross}
\end{equation}
For hadron beams the EPA has better accuracy than the well known Weizs\"{a}cker-Williams
 approximation. 
\par A similar mechanism of two-photon production occurs during heavy ion collisions where the coherent radiation of photons are enhanced by the number of protons, $Z$, in the nuclei \cite{Drees:1989vq,Papageorgiu:1995eg}.
In this case, however, one has to impose the restriction that the nuclei have an impact parameter
greater then $b_{min}=2R$, where $R$ is the nuclear radius. This restriction  reduces the cross section substantially~\cite{Cahn:1990jk}.

\subsubsection{Standard model two-photon processes}
Two photon induced exclusive processes are selected using \nflux=12, 13, 14, 15, 16 and \typepr='EXC', \typint='QED'.

The following processes are available:

\begin{center}
\begin{tabular}{|r|l|}
\hline
\textbf{IPROC} & \textbf{Description} \\
\hline
        16006 & $\gamma\gamma\rightarrow ll$ \\
        16010 & $\gamma\gamma\rightarrow W^+W^-$\\
\hline
\end{tabular}
\end{center}
Note that details of the generation such as kinematic ranges are steered
by parameters defined in \herwig{} for the particular processes numbers \iproc.

\subsubsection{Beyond standard model two-photon processes}
The following beyond Standard Model processes are currently supported:
\begin{center}
\begin{tabular}{|r|r|l|}
\hline
\textbf{IPROC} & \textbf{AAANOM} &\textbf{Description} \\
\hline
        16010 & 2& $\gamma\gamma\rightarrow W^+W^-$ beyond SM \\
        16015 & 3& $\gamma\gamma\rightarrow ZZ$  beyond SM \\
\hline
\end{tabular}
\end{center}

Two-photon events can be used for studies of a photon coupling to other gauge bosons. \fpmc{} was interfaced with a code produced by \comphep{} providing amplitudes of di-boson production $\gamma\gamma\rightarrow WW$ induced by effective Lagrangians
due to anomalous couplings between $\gamma$ and $W$ or $Z$.
In particular, in \fpmc{} one can study the C,P-parity conserving Triple Gauge Coupling
 (TGC) and Quartic Gauge Coupling (QGC) coupling  parameterized by two anomalous 
parameters and four anomalous quartic parameters (see App.~\ref{app:anomlag} for complete form of the Lagrangians). 
The effect of the anomalous Lagrangian can be regulated with a form factor in a dipole form $\lambda\rightarrow\lambda/(1+ (s_{\gamma\gamma}/\Lambda_\mathrm{cutoff})^2)^2$, 
where $s_{\gamma\gamma}$ is the invariant photon-photon center-of-mass and $\Lambda_\mathrm{cutoff}$ is a typical scale of the new physics. Parameters are:
\begin{center}
\begin{tabular}{|r|c|c|p{8cm}|}
\hline
\textbf{Parameter} & \textbf{Type} & \textbf{Default} & \textbf{Description}\\
\hline
DKAPPA    & float          & -    & trilinear couping $\dkap$ (Eq.~\refeq{eq:TGClag} in App.~\ref{app:anomlag})\\
LAMBDA    & [GeV$^{2}$]    & -    & trilinear couping $\lam$  \\
A0W       & [GeV$^{-2}$]   & -    & quartic coupling $\aOw/\Lambda$ (Eq.~\refeq{eq:QGClag} in App.~\ref{app:anomlag})\\
ACW       & [GeV$^{-2}$]   & -    & quartic coupling $\aCw/\Lambda$  \\
A0Z       & [GeV$^{-2}$]   & -    & quartic coupling $\aOz/\Lambda$  \\
ACZ       & [GeV$^{-2}$]   & -    & quartic coupling $\aCz/\Lambda$  \\
ANOMCUTOFF & [GeV$^{2}$], -1& -1(OFF)   & form factor $\Lambda_\mathrm{cutoff}$ (see text)  \\
\hline
\end{tabular}
\end{center}

\subsection{Central exclusive QCD production}

Central exclusive processes can occur also via the strong interactions. In
analogy to the QED case, one often talks about the Pomeron exchange and
$\mathbb{P}\mathbb{P}\rightarrow X$ sub-process. However, in reality the
calculations are performed in perturbative QCD as a two-gluon exchange. Such a
process consists of a hard sub-process $gg\rightarrow X$ and an additional gluon
that screens the color. The screening gluon makes the exchange a color singlet.

There are several models of such interactions on the market~\cite{Khoze:2001xm,Cudell:2008gv,Maciula:2010tv}, differing in the treatment of the proton structure, the
virtual corrections and the approximations. All models consist of soft
re-scattering corrections (the Rapidity Gap Survival Probability) and some of
them~\cite{Khoze:2001xm,Ryskin:2009tj,Ryskin:2009tk,Martin:2010rn,Frankfurt:2006jp,Gotsman:2005wa} introduce non-perturbative elements to the
calculation.

Currently there are two models of central exclusive production implemented into the FPMC: the KMR (Durham) and the CHIDe (Liege) model.

\subsubsection{The KMR (Durham) Model}

The details of the models can be found in \cite{Khoze:2001xm}. The model can be chosen
by setting \nflux=16 and \typepr='EXC', \typint='QCD'. The available sub-processes are:
\begin{center}
\begin{tabular}{|r|l|}
\hline
\textbf{IPROC} & \textbf{Description} \\
\hline
16013 & \cep{gg/q\bar q}\\
19999 & \cep{H} \\
\hline
\end{tabular}
\end{center}
The following parameters of the KMR model can be modified in the steering file:
%
\begin{center}
\begin{tabular}{|r|c|c|p{8cm}|}
\hline
\textbf{Parameter} & \textbf{Type} & \textbf{Default} & \textbf{Description}\\
\hline
Q2CUT & [GeV$^2$]      & 2.0  & Lower limit of luminosity integration \\
SURV  & \textit{float} & 0.03 & Rapidity Gap Survival Probability \\
SCALE & \textit{float} & 1.0  & Scales the upper limit in Sudakov formfactor \\
DELTA & {1,2}          & 2    & Definition of $\Delta$ \\ 
\hline
\end{tabular}
\end{center}

\subsubsection{The CHIDe (Liege) Model}

The details of the CHIDe model can be found in \cite{Cudell:2008gv} and it is set by
NFLUX=18 nd \typepr='EXC', \typint='QCD'. The available subprocesses are:
\begin{center}
\begin{tabular}{|r|l|}
\hline
\textbf{IPROC} & \textbf{Description} \\
\hline
16012 & \cep{gg} \\
19999 & \cep{H} \\
\hline
\end{tabular}
\end{center}
The following parameters of the CHIDe model can be modified in the steering file:
\begin{center}
\begin{tabular}{|r|c|c|p{8cm}|}
\hline
\textbf{Parameter} & \textbf{Type} & \textbf{Default} & \textbf{Description}\\
\hline
IGLU   & \{1,2,3,4\}    & 4    & Different impact factor (parameterization of
                                 the gluon density) \\
LSCALE & \textit{float} & 1.0  & Scales the lower limit in Sudakov formfactor\\
USCALE & \textit{float} & 1.0  & Scales the upper limit in Sudakov formfactor\\
SURV   & \textit{float} & 0.03 & Rapidity Gap Survival Probability \\ 
\hline
\end{tabular}
\end{center}
%
%

\subsection{Inclusive hard diffraction - Pomeron/Reggeon exchange}
\label{ExcPomMod}

The implementation of inclusive hard diffractive processes follows the Ingelman-Schlein model of diffraction~\cite{Ingelman:1984ns}. The structure of the proton
in events with rapidity gaps is modeled by a color singlet Pomeron/Reggeon exchange. We 
distinguish single diffractive  events with single pomeron exchange
\begin{equation}
  p p \rightarrow p \oplus X + \mathrm{Pomeron\ remnants} + \mathrm{proton\ remnants}
\end{equation}
and double Pomeron exchange 
\begin{equation}
  p p \rightarrow p \oplus X \oplus p + \mathrm{Pomeron\ remnants},
\end{equation}
where $ \oplus$ indicate the presence of the pseudo-rapidity gap.

The diffractive cross section  of single diffractive dissociation (SD) is calculated as a convolution of the diffractive
structure function and the partonic sub-process cross section 

\be
\d\sigma^{pp\rightarrow pX}=\sum_i\int f^D_i(x_i, \mu^2, \xi, t) f_j(x_j, \mu^2)d\sigma^{i,j\rightarrow X}_\mathrm{sub}(x_i, x_j, \mu^2)\,\d x_i\,\d x_j\,\d \xi\,\d t
\ee
where $x_i,\,x_j$ are the Bjorken-$x$  of the parton coming from Pomeron and proton respectively, and $\mu^2$ are the renormalization and factorization scales, set equal in the formula above for clarity. Calculation of the double pomeron exchange (DPE) is done along the same line
 substituting the non-diffractive $f_j$ distribution by the diffractive $f^{D}_j$ one.
\par 
Whereas this factorization has been proven to be valid in the $ep$
 collider, there is an additional suppressing factor for hadron-hadron collisions, so called survival probability factor. 
At hadron-hadron collider, this suppression arises from soft interaction between the incoming/outgoing
 hadrons leading to the proton break-up and loss of the event diffractive signature. The factor is believed
to be weakly dependent on the particular process. 
\par 
Measurements at HERA showed that the hadron structure in diffraction can be described in terms
of the parton density functions (PDF) in the same way as in non-diffractive case, however in redefined kinematics dependent on the pomeron longitudinal momentum fraction $\xi$. This in diffraction commonly used variable is related to the mass of the created diffractive system as $\xi=M^2_X/s$, $s$ being the center-of-mass energy of the proton-proton collision. For the structure function, the following factorization has been observed
\be
f^D_i(x, \mu^2, \xi, t)=f_{\pom/p}(\xi,t)\cdot f_{i/\pom}(\beta=x/\xi,\mu^2).
\label{eq:sm:protonvertex}
\ee 
 The production of large diffractive 
masses is suppressed by the pomeron flux $f_{\pom/p}(\xi,t)$ which 
approximately behaves as $\sim1/\xi$ for not to high diffractive masses 
(and above a resonance region $>\sim1$ GeV).
 The normalization of the fluxes is conventionally fixed at $\xi_\pom=0.003$ such that
\be
\xi_\pom \int_{t_\mathrm{cut}}^{t_\mathrm{min}}f_{\pom/p}\,\d t =1, 
\ee
where $|t_\mathrm{min}|\simeq m_p^2x_\pom^2/(1-x_\pom)$ is the minimum kinematically accessible value of $|t|$, $m_p$ is the proton mass and $|t_\mathrm{cut}|=1.0\GeV^2$~\cite{Aktas:2006hy}. Note that the same structure function can be assigned to a Reggeon, but the Reggeon contribution is expected to be small at high energy.
\par 
The cross section for inclusive DPE processes can schematically be expressed as
\begin{equation}
	d\sigma^{\mathrm{pp\rightarrow ppX}} = \sum_{i,j}\int dx_i dx_j \, d\xi_i d\xi_j \, F_{\pom/p}(\xi_i) F_{\pom/p}(\xi_j) \, f_{i/\pom}(x_i,\mu^2) f_{j/\pom}(x_j,\mu^2) \, d\hat{\sigma}(ij\rightarrow X),
\end{equation}
where $F_{\pom/p}(\xi)$ is the pomeron flux and $f_{i/\pom}(x,\mu^2)$ parton density in the Pomeron described above.

\subsubsection{Implementation}
The only difference in the implementation of the single diffractive and double pomeron exchange processes
 with respect to non-diffractive production already present in \herwig{} is the substitution of the 
proton density function by the proton diffractive structure function in Eq.~\refeq{eq:sm:protonvertex}.
Some of the diffractive processes which are of interest at the LHC are listed below, but the user
can in principle choose other hard processes defined in \herwig{} and run them in the inclusive diffractive mode. 
The inclusive mode is selected by \nflux=9, 10 (Reggeon or Pomeron exchange, see Table~\ref{tab:nflux}) together with \typepr='INC' 
and \typint='QCD'.
\begin{center}
\begin{tabular}{|r|l|}
\hline
\multicolumn{2}{|c|}{\bf Single diffraction/Double pomeron exchange} \\
\hline
\hline
\textbf{IPROC}  &\textbf{Description} \\
\hline
11300 & $q\bar{q}\rightarrow Z/\gamma\rightarrow q'\bar{q}'$ \\
11350 & $q\bar{q}\rightarrow Z/\gamma\rightarrow l\bar{l}$ \\
11399 & $q\bar{q}\rightarrow Z/\gamma\rightarrow$ any\\
      11400 & $q\bar{q}\rightarrow W^{\pm}\rightarrow l\nu_l$ \\
      11450 & $q\bar{q}\rightarrow W^{\pm}\rightarrow l\nu_l$ \\
      11499 & $q\bar{q}\rightarrow W^{\pm}\rightarrow$ any\\
\hline 
11500 & QCD $2\rightarrow2$ parton scattering\\
11700 & QCD heavy quark production \\
\hline
12200 & QCD direct photon pair production\\
\hline
16010 & $W^+W^-$\\
\hline
\end{tabular}
\end{center}

\subsubsection{PDF}
Parton densities in the Pomeron are selected with the {\tt IFIT} parameter. H1 fits required 
both the Pomeron and Reggeon exchanges to be included to describe the data 
 \cite{Aktas:2006hy}, though the Reggeon parton density was assumed to be the 
PDF of a pion. At high energy, the Pomeron contributions dominate.  The production
 through Pomeron/Reggeons only can be selected via  \nflux=9, 10.
  \begin{center}
\begin{tabular}{|c|l|c|}
    	\hline
			\textbf{IFIT} & \textbf{PDF set} & Reference\\ \hline  \hline
         101& Official H1 fit B (default) & \cite{Aktas:2006hy} \\
         100& Official H1 fit A & \cite{Aktas:2006hy}\\ 
\hline
		\end{tabular}
  \end{center}

\subsection{Rescattering corrections}
The Factorization theorem which, in standard non-diffractive production, allows to use PDF measured
in one processes for other theoretical predictions does not hold in diffractive or exclusive
processes in hadron-hadron colliders. The factorization breaking suppression is energy dependent; it was measured to be about 0.1 at the Tevatron and
is expected to be about 0.03 at the LHC~\cite{Khoze:2001xm,Ryskin:2009tj,Ryskin:2009tk,Martin:2010rn,Frankfurt:2006jp,Gotsman:2005wa}. 
This correction slightly depends on the type and kinematic
of the process. They may modify the angular distribution of scattered final state protons.
Several models are therefore included in \fpmc: KMR low mass diffractive
model and Effective opacity model studied at~\cite{Kupco:2004fw}. The options
are
\begin{center}
\begin{tabular}{|r|l|}
\hline
{\bf ISOFTM} & {\bf Description }  \\
\hline
0 & No correction\\
         1 & Constant factor~\cite{Khoze:2001xm}\\
         2 & KMR low mass diffractive model~\cite{Kupco:2004fw}\\
         3 & Effective opacity model \cite{Kupco:2004fw}\\
\hline
\end{tabular}
\end{center}


\section{Conclusions / perspectives}
\label{sec:conclusion}
In a relatively simple way, routines of \herwig{} are replaced in \fpmc{} to implement wide range 
of processes with leading intact protons. These can represent signal, background or both 
at the same time for particular data analysis,
 and it is therefore important to accommodate all into one framework with the same hadronization model. 
\par
Some of these like the Ingelman-Schlein have been already implemented in more modern version of
\herwig{} (\herwigcc), but for others like the exclusive KMR and CHIDe model, \fpmc{} is the only 
generator were both are implemented and can thus be clearly compared to each other for the 
exclusive Higgs or di-jet productions for instance.
\par
We should not forget to mention the importance of two-photon exclusive processes, especially 
the di-boson anomalous production which when observed can give a strong evidence 
for beyond standard electroweak symmetry breaking. 
\par
Several other processes might be worth of implementing in the future, namely the
exclusive QCD production of di-photons or $\chi_c$, $\chi_b$ or other models 
for anomalous productions in photon-photon interactions as they will be studied
at the LHC.


\bibliographystyle{atlasnote}
\bibliography{fpmc_manual_v1.0}

\providecommand{\href}[2]{#2}\begingroup\raggedright\begin{thebibliography}{10}

\bibitem{Ingelman:1984ns}
G.~Ingelman and P.~E. Schlein, {\em {Jet Structure in High Mass Diffractive
  Scattering}\/},
\href{http://dx.doi.org/10.1016/0370-2693(85)91181-5}{Phys. Lett. {\bf B152}
  (1985)  256}.

\bibitem{Cahn:1990jk}
R.~N. Cahn and J.~D. Jackson, {\em {Realistic equivalent photon yields in heavy
  ion collisions}\/},
\href{http://dx.doi.org/10.1103/PhysRevD.42.3690}{Phys. Rev. {\bf D42} (1990)
  3690--3695}.

\bibitem{Drees:1989vq}
M.~Drees, J.~R. Ellis, and D.~Zeppenfeld, {\em {Can one detect an intermediate
  mass higgs boson in heavy ion collisions?}\/},
\href{http://dx.doi.org/10.1016/0370-2693(89)91632-8}{Phys. Lett. {\bf B223}
  (1989)  454}.

\bibitem{Papageorgiu:1995eg}
E.~Papageorgiu, {\em {An Intermediate mass Higgs boson in two photon coherent
  processes at the LHC}\/},
  \href{http://dx.doi.org/10.1016/0370-2693(95)00434-M}{Phys. Lett. {\bf B352}
  (1995)  394--399},
\href{http://arxiv.org/abs/hep-ph/9503372}{{\tt arXiv:hep-ph/9503372}}.

\bibitem{Budnev:1974de}
V.~M. Budnev, I.~F. Ginzburg, G.~V. Meledin, and V.~G. Serbo, {\em {The Two
  photon particle production mechanism. Physical problems. Applications.
  Equivalent photon approximation}\/},
\href{http://dx.doi.org/10.1016/0370-1573(75)90009-5}{Phys. Rept. {\bf 15}
  (1975)  181--281}.

\bibitem{Khoze:2001xm}
V.~A. Khoze, A.~D. Martin, and M.~G. Ryskin, {\em {Prospects for new physics
  observations in diffractive processes at the LHC and Tevatron}\/},
  \href{http://dx.doi.org/10.1007/s100520100884}{Eur. Phys. J. {\bf C23} (2002)
   311--327},
\href{http://arxiv.org/abs/hep-ph/0111078}{{\tt arXiv:hep-ph/0111078}}.

\bibitem{Cudell:2008gv}
J.~R. Cudell, A.~Dechambre, O.~F. Hernandez, and I.~P. Ivanov, {\em {Central
  exclusive production of dijets at hadronic colliders}\/},
  \href{http://dx.doi.org/10.1140/epjc/s10052-009-0994-2}{Eur. Phys. J. {\bf
  C61} (2009)  369--390},
\href{http://arxiv.org/abs/0807.0600}{{\tt arXiv:0807.0600 [hep-ph]}}.

\bibitem{Corcella:2002jc}
G.~Corcella et al., {\em {HERWIG 6.5 release note}\/},
\href{http://arxiv.org/abs/hep-ph/0210213}{{\tt arXiv:hep-ph/0210213}}.

\bibitem{fpmc}
{\em {FPMC home page, \url{www.cern.ch/fpmc}}\/},  2011.

\bibitem{Maciula:2010tv}
R.~Maciula, R.~Pasechnik, and A.~Szczurek, {\em {Central exclusive
  quark-antiquark dijet and Standard Model Higgs boson production in
  proton-(anti)proton collisions}\/},
\href{http://arxiv.org/abs/1011.5842}{{\tt arXiv:1011.5842 [hep-ph]}}.

\bibitem{pompyt}
{\em {POMPYT generator for diffraction in Pythia
  \url{http://www3.tsl.uu.se/thep/pompyt/}}\/},  1997.

\bibitem{Cox:2000jt}
B.~E. Cox and J.~R. Forshaw, {\em {POMWIG: HERWIG for diffractive
  interactions}\/},
  \href{http://dx.doi.org/10.1016/S0010-4655(01)00467-2}{Comput. Phys. Commun.
  {\bf 144} (2002)  104--110},
\href{http://arxiv.org/abs/hep-ph/0010303}{{\tt arXiv:hep-ph/0010303}}.

\bibitem{Boonekamp:2003ie}
M.~Boonekamp and T.~Kucs, {\em {Pomwig v2.0: Updates for double
  diffraction}\/},  \href{http://dx.doi.org/10.1016/j.cpc.2005.01.002}{Comput.
  Phys. Commun. {\bf 167} (2005)  217},
\href{http://arxiv.org/abs/hep-ph/0312273}{{\tt arXiv:hep-ph/0312273}}.

\bibitem{Monk:2005ji}
J.~Monk and A.~Pilkington, {\em {ExHuME: A Monte Carlo event generator for
  exclusive diffraction}\/},
  \href{http://dx.doi.org/10.1016/j.cpc.2006.04.005}{Comput. Phys. Commun. {\bf
  175} (2006)  232--239},
\href{http://arxiv.org/abs/hep-ph/0502077}{{\tt arXiv:hep-ph/0502077}}.

\bibitem{Baranov:1991yq}
S.~P. Baranov, O.~Duenger, H.~Shooshtari, and J.~A.~M. Vermaseren, {\em {LPAIR:
  A generator for lepton pair production}\/}, . In *Hamburg 1991, Proceedings,
  Physics at HERA, vol. 3* 1478-1482. (see HIGH ENERGY PHYSICS INDEX 30 (1992)
  No. 12988).

\bibitem{Ryskin:2009tj}
M.~G. Ryskin, A.~D. Martin, and V.~A. Khoze, {\em {Soft processes at the LHC,
  I: Multi-component model}\/},
  \href{http://dx.doi.org/10.1140/epjc/s10052-009-0889-2}{Eur. Phys. J. {\bf
  C60} (2009)  249--264},
\href{http://arxiv.org/abs/0812.2407}{{\tt arXiv:0812.2407 [hep-ph]}}.

\bibitem{Ryskin:2009tk}
M.~G. Ryskin, A.~D. Martin, and V.~A. Khoze, {\em {Soft processes at the LHC,
  II: Soft-hard factorization breaking and gap survival}\/},
  \href{http://dx.doi.org/10.1140/epjc/s10052-009-0890-9}{Eur. Phys. J. {\bf
  C60} (2009)  265--272},
\href{http://arxiv.org/abs/0812.2413}{{\tt arXiv:0812.2413 [hep-ph]}}.

\bibitem{Martin:2010rn}
A.~D. Martin, M.~G. Ryskin, and V.~A. Khoze, {\em {Towards a model which merges
  soft and hard high-energy pp interactions}\/},
\href{http://arxiv.org/abs/1011.0287}{{\tt arXiv:1011.0287 [hep-ph]}}.

\bibitem{Frankfurt:2006jp}
L.~Frankfurt, C.~E. Hyde, M.~Strikman, and C.~Weiss, {\em {Generalized parton
  distributions and rapidity gap survival in exclusive diffractive $p p$
  scattering}\/},  \href{http://dx.doi.org/10.1103/PhysRevD.75.054009}{Phys.
  Rev. {\bf D75} (2007)  054009},
\href{http://arxiv.org/abs/hep-ph/0608271}{{\tt arXiv:hep-ph/0608271}}.

\bibitem{Gotsman:2005wa}
E.~Gotsman, H.~Kowalski, E.~Levin, U.~Maor, and A.~Prygarin, {\em {Survival
  probability for diffractive di-jet production at the LHC}\/},
  \href{http://dx.doi.org/10.1140/epjc/s2006-02600-1}{Eur. Phys. J. {\bf C47}
  (2006)  655--669},
\href{http://arxiv.org/abs/hep-ph/0512254}{{\tt arXiv:hep-ph/0512254}}.

\bibitem{Aktas:2006hy}
{H1} Collaboration, A.~Aktas et al., {\em {Measurement and QCD analysis of the
  diffractive deep- inelastic scattering cross-section at HERA}\/},
  \href{http://dx.doi.org/10.1140/epjc/s10052-006-0035-3}{Eur. Phys. J. {\bf
  C48} (2006)  715--748},
\href{http://arxiv.org/abs/hep-ex/0606004}{{\tt arXiv:hep-ex/0606004}}.

\bibitem{Kupco:2004fw}
A.~Kupco, C.~Royon, and R.~B. Peschanski, {\em {Decisive test for the pomeron
  at Tevatron}\/},
  \href{http://dx.doi.org/10.1016/j.physletb.2004.11.072}{Phys. Lett. {\bf
  B606} (2005)  139--144},
\href{http://arxiv.org/abs/hep-ph/0407222}{{\tt arXiv:hep-ph/0407222}}.

\bibitem{Boos:2004kh}
{CompHEP} Collaboration, E.~Boos et al., {\em {CompHEP 4.4: Automatic
  computations from Lagrangians to events}\/},
  \href{http://dx.doi.org/10.1016/j.nima.2004.07.096}{Nucl. Instrum. Meth. {\bf
  A534} (2004)  250--259},
\href{http://arxiv.org/abs/hep-ph/0403113}{{\tt arXiv:hep-ph/0403113}}.

\bibitem{Belanger:1992qh}
G.~Belanger and F.~Boudjema, {\em {Probing quartic couplings of weak bosons
  through three vectors production at a 500-GeV NLC}\/},
\href{http://dx.doi.org/10.1016/0370-2693(92)91978-I}{Phys. Lett. {\bf B288}
  (1992)  201--209}.

\bibitem{Hagiwara:1986vm}
K.~Hagiwara, R.~D. Peccei, D.~Zeppenfeld, and K.~Hikasa, {\em {Probing the Weak
  Boson Sector in e+ e- $\to$ W+ W-}\/},
\href{http://dx.doi.org/10.1016/0550-3213(87)90685-7}{Nucl. Phys. {\bf B282}
  (1987)  253}.

\end{thebibliography}\endgroup

\appendix

\section{Manual} \label{appa}
\subsection{Usage}

FPMC runs as a standalone program. It requires installed CERN libraries, fortran compiler (gfortran/g77), 
and c++ compiler.  The program is compiled with 
\begin{verbatim}
make 
\end{verbatim}

Two binary will compile: {\verb module } and {\verb module_reco }. Both modules are designed to be run with a data card that changes the default parameters. In the latter, a simple jet reconstruction algorithm is run and hadron level final states are saved in an ntuple. 
Several data cards are provided in {\tt Datacards/} directory for some standard processes of interest.
 For example, the $WW$ production via $\gamma\gamma$ fusion is executed by 
\begin{verbatim}
./module < Datacards/dataQEDWW
\end{verbatim}

\subsection{Parameters}
The main parameters related to the processes with leading intact protons are summarized in Table~\ref{tab:parameters}.
\begin{table}[htb!]
 \begin{center}
 \caption{Parameters which can be set through data cards.}
 \vspace{0.25cm}
  \begin{tabular}{|r|l|c|} 
    \hline
     \textbf{Parameter} & \textbf{Description} & \textbf{Default}  \\ \hline \hline 
			 TYPEPR	&		Select  exclusive 'EXC' or inclusive 'INC' production		&	'EXC'		\\ \hline
			 TYPINT	&		Switch between QED and QCD process		                  &	'QED'		\\ \hline
			 NFLUX	&		Select flux		                                          &	15		   \\ \hline
          IPROC	&	   Type of process to generate		                        &  11500    \\ \hline
	       MAXEV	&	   Number of events to generate		                        &  100	   \\ \hline
			 PBEAM1&	   Type of beam 1 particle 		                           &  E+   	   \\ \hline
			 PBEAM2&     Type of beam 2	particle       	                        &	E+	      \\ \hline 
          ECMS    &     CMS energy (in GeV)                                      &  14000    \\ \hline 
          HMASS   &     Higgs mass (GeV)                                         & 115       \\ \hline
			 PTMIN	&     Minimum $p_T$ in hadronic jet production                 &  0        \\ \hline
          PTMAX	&     Maximum $p_T$ in hadronic jet production                 &  $10^8$   \\ \hline
			 YJMIN	&     Minimum jet rapidity		                                 &  -6       \\ \hline
			 YJMAX	&     Maximum jet rapidity		                                 &  +6       \\ \hline
			 EEMIN	&	   Minimum dilepton mass in Drell-Yan	                     &	$10.0$	\\ \hline
			 EEMAX	&	   Maximum dilepton mass in Drell-Yan		                  &	$10^8$	\\ \hline
			 IFITPDF &		Diffractive PDF                                        	&	100		   \\ \hline
          NTNAME  &     Output ntuple name                                       &  'tmpntuple.ntp' \\ \hline
			 ISOFTM	&		Soft correction	                                       &	1	      \\ \hline
			 IAION		&		Atomic number of colliding nuclei		                  &  1        \\ \hline
			 IZION		&		Proton number of colliding nuclei		                  &  1        \\ \hline
          NRN1    &     1. random number generator initial seed                  &           \\ \hline
          NRN2    &     2. random number generator initial seed                  &           \\ \hline
			 Q2WWMN  &     Minimum momentum transfer ($Q^2 = |t|$)		            &	0	      \\ \hline
			 Q2WWMX	&	   Maximum momentum transfer ($Q^2 = |t|$)		            &	4	      \\ \hline
			 YWWMIN	&	   Minimum beam momentum loss ($\xi_{min}$)	               &	0	      \\ \hline
			 YWWMAX	&	   Maximum beam momentum loss ($\xi_{max}$)		            &	0.1	   \\ \hline
\hline
\end{tabular}
\end{center}
\label{tab:parameters}
\end{table}

\newpage

\subsubsection{Processes with two leading protons}
All processes in this section assume initial generator setup of two electron beams \textbf{PBEAM1}='E+' and \textbf{PBEAM2}='E+'. They are then internally converted to diffractive protons and appropriate PDF is called in case of double pomeron exchange. In case of exclusive 
production, the appropriate model for gluon-gluon exchange is used.  

	\begin{center}
  \begin{tabular*}{\textwidth}{@{\extracolsep{\fill}}|l|c|c|c|c|} 
    \hline
    \hline
  \multicolumn{5}{|c|}{\bf  Higgs processes } \\
\hline
\hline
   \parbox{\colfirst}{  \textbf{Process}}   & \textbf{IPROC} & \textbf{TYPEPR} & \textbf{TYPINC} & \textbf{NFLUX} \\ \hline  \hline
			Incl. $H$ & 11600+ID & INC & QCD &  9,10,11 \\ \hline 
			Excl. $H$ & 19900+ID & EXC & QCD &  16 \\ \hline 
			Excl. $H$ & 19900+ID & EXC & QED &  12,13,14,15 \\ \hline
			\multicolumn{2}{|r}{ID} & \multicolumn{1}{@{}l}{ = 1,\dots,6} &\multicolumn{2}{l|}{ $H\rightarrow q\bar{q}$ (resp. d,u,s,c,b,t) \hspace{2.5cm}} \\ 
			\multicolumn{2}{|r}{ID} & \multicolumn{1}{@{}l}{ = 7,8,9} & \multicolumn{2}{l|}{$H\rightarrow l^+ l^-$ (resp. $e^+ e^-,\mu^+ \mu^-, \tau^+ \tau^-$)} \\ 
			\multicolumn{2}{|r}{ID} & \multicolumn{1}{@{}l}{ = 10,11} & \multicolumn{2}{l|}{$H\rightarrow W^+ W^-,ZZ$} \\ 
			\multicolumn{2}{|r}{ID} & \multicolumn{1}{@{}l}{ = 99} & \multicolumn{2}{l|}{all decay modes}\\ \hline
	\end{tabular*}
	\end{center}
	\begin{center}
  \begin{tabular*}{\textwidth}{@{\extracolsep{\fill}}|l|c|c|c|c|} 
    \hline
    \hline
  \multicolumn{5}{|c|}{\bf  Dijet processes } \\
\hline
\hline
      \parbox{\colfirst}{  \textbf{Process}}  & \textbf{IPROC} & \textbf{TYPEPR} & \textbf{TYPINC} & \textbf{NFLUX} 
		\\ \hline  \hline 
			Incl. dijets   & 11500 &  INC & QCD &  9,10,11 \\ 
			\hline
			Incl. heavy $q\bar{q}$  & 11700+ID & INC & QCD &  9,10,11 \\ 
			\hline
            \multicolumn{3}{|l}{\hspaceID ID = 1,\dots,6} & \multicolumn{2}{l|}{$gg\rightarrow q\bar{q}$ (resp. d,u,s,c,b,t)}\\ \hline	
			Excl. $q\bar{q}$  & 16000+ID & EXC & QCD &  16, 17 \\ 
    \hline
	\multicolumn{3}{|l}{\hspaceID 	ID = 1,\dots,6} & \multicolumn{2}{l|}{$gg\rightarrow q\bar{q}$ (resp. d,u,s,c,b,t)}\\ 
	\multicolumn{3}{|l}{\hspaceID ID = 11} & \multicolumn{2}{l|}{$gg\rightarrow q\bar{q}$ (all flavours)} \\
	\multicolumn{3}{|l}{\hspaceID ID = 13} & \multicolumn{2}{l|}{$gg\rightarrow gg + q\bar{q}$ (all flavours)} \\ 
    \hline 
			Excl. dijets & 16000+ID& EXC & QED &  16, 17 \\ 
    \hline
	\multicolumn{3}{|l}{\hspaceID  ID	= 1,\dots,6} & \multicolumn{2}{l|}{$gg\rightarrow q\bar{q}$ (resp. d,u,s,c,b,t) \hspace{2.5cm}}\\ \hline	
    \multicolumn{3}{|l}{\hspaceID  ID		= 7,8,9} & \multicolumn{2}{l|}{$\gamma\gamma\rightarrow l^+ l^-$ (resp. $e^+ e^-,\mu^+ \mu^-, \tau^+ \tau^-$)} \\ \hline
	\end{tabular*}
	\end{center}

	\begin{center}
  \begin{tabular*}{\textwidth}{@{\extracolsep{\fill}}|l|c|c|c|c|} 
\hline
    \hline
  \multicolumn{5}{|c|}{\bf   $W^+W^-$, photon and lepton pairs  } \\
\hline
    \hline
     \parbox{\colfirst}{  \textbf{Process}}  & \textbf{IPROC}  & \textbf{TYPEPR} & \textbf{TYPINC} & \textbf{NFLUX} \\ \hline  \hline
			Incl. $W^+W^-$ & 12800 & INC & QCD & 9,10,11 \\ \hline
			Excl. $W^+W^-$ & 16010 & EXC & QED & 15 \\ \hline
			Incl. $\gamma\gamma$  & 12200& INC & QCD &  9,10,11 \\ \hline
			Excl. $\gamma\gamma$  & 19800 & INC & QCD &  16 \\ \hline
			Excl. $\gamma\gamma$  & 19800 & EXC & QED &  12,13,14,15 \\ \hline
			Excl. $ll$  & 16006+IL & EXC & QED &  12,13,14,15 \\ \hline
			Incl. $ll$ & 11350  & INC & QCD &  9,10,11 \\ \hline
			Incl. $ll$ & 11350+IL & INC & QCD &  9,10,11 \\ \hline
	\multicolumn{3}{|l}{\hspaceID  IL = 0,1,2,3} & \multicolumn{2}{l|}{(resp. all families, $e,\mu,\tau$)} \\ \hline
	\end{tabular*}
	\end{center}

\subsubsection{Single diffraction}
All processes in this section assume initial setup with one electron beam \textbf{PBEAM1}='P' and \textbf{PBEAM2}='E+'. The electron is then internally converted to a diffractive proton. 

	\begin{center}
  \begin{tabular*}{\textwidth}{@{\extracolsep{\fill}}|l|l|c|c|c|}

    \hline
    \hline
  \multicolumn{5}{|c|}{\bf  Single diffraction } \\
\hline
\hline
      \parbox{\colfirst}{  \textbf{Process}}  & \textbf{IPROC} & \textbf{TYPEPR} & \textbf{TYPINC} & \textbf{NFLUX} \\ \hline  \hline
			Incl. SD $Z\rightarrow q\bar{q}$ & 11300+IQ &INC & QCD &  9,10,11 \\ \hline
	\multicolumn{3}{|l}{\hspaceID  IQ = 0,1,\dots,6} & \multicolumn{2}{l|}{(resp. all flavours, d,u,s,c,b,t)} \\ \hline
			Incl. SD $Z\rightarrow l\bar{l}$ & 11350+IL &INC & QCD &  9,10,11 \\ \hline
			Incl. SD $Z\rightarrow$ any & 11399 &INC & QCD &  9,10,11 \\ \hline

			Incl. SD $W^{\pm}\rightarrow qq'$ & 11400+IQ &INC & QCD &  9,10,11 \\ \hline
	\multicolumn{3}{|l}{\hspaceID  IQ = 0,1,\dots,6} & \multicolumn{2}{l|}{(resp. all flavours, d,u,s,c,b,t)} \\ \hline
			Incl. SD $W^{\pm}\rightarrow l\nu$ & 11450+IL &INC & QCD &  9,10,11 \\ \hline
	\multicolumn{3}{|l}{\hspaceID  IL = 0,1,\dots,3} & \multicolumn{2}{l|}{( $l=$ all families, $e,\mu,\tau$)} \\ \hline
			Incl. SD $W^{\pm}\rightarrow$ any & 11499 &INC & QCD &  9,10,11 \\ \hline
			Incl. SD dijets & 11500 & INC & QCD &  9,10,11 \\ \hline
	\end{tabular*}
	\end{center}

\section{Anomalous $\gamma\gamma\rightarrow WW$ coupling} \label{app:anomlag}
\par The amplitude allowing studies of the anomalous coupling of the photon to 
$W$ boson was generated by \comphep~\cite{Boos:2004kh}. 
For triple gauge boson coupling the following effective Lagrangian was assumed
 \begin{equation}
   \mathcal{L}/ig_{WW\gamma}=(W^{\dagger}_{\mu\nu}W^{\mu}A^{\nu}-W_{\mu\nu}W^{\dagger\mu}A^{\nu})
   +(1+{\Delta\kappa^{\gamma}})W_{\mu}^{\dagger}W_{\nu}A^{\mu\nu}+\frac{{\lambda^{\gamma}}}{M_W^2}W^{\dagger}_{\rho\mu}
   W^{\mu}_{\phantom{\mu}\nu}A^{\nu\rho}.
\label{eq:TGClag}
\end{equation}
There, $g_{WW\gamma}=-e$ is the standard $\wwgamma$ coupling in the SM and the double-indexed terms are $ V_{\mu\nu}\equiv\partial_{\mu} V_{\nu}-\partial_{\nu}V_{\mu}\nonumber$, for $V^{\mu}=W^{\mu},A^{\mu}$. The Lagrangian considered contains terms which conserve
C,P-parity separately only. 

\par On the other hand, the effective quartic boson coupling was taken to be parameterized by four anomalous parameters $\aOw, \aOz, \aCw$, and $\aCz$

\begin{eqnarray}
\mathcal{L}_6^0 & = & \frac{-e^2}{8} \frac{\aOw}{\Lambda^2} F_{\mu\nu} F^{\mu\nu} W^{+\alpha} W^-_\alpha \label{eq:QGClag}\\
      &&- \frac{e^2}{16\cos^2 \Theta_W} \frac{\aOz}{\Lambda^2} F_{\mu\nu} F^{\mu\nu} Z^\alpha Z_\alpha, \nonumber \\
      \mathcal{L}_6^C & = & \frac{-e^2}{16} \frac{\aCw}{\Lambda^2} F_{\mu\alpha} F^{\mu\beta} (W^{+\alpha} W^-_\beta + W^{-\alpha} W^+_\beta) \nonumber \\
      &&- \frac{e^2}{16\cos^2 \Theta_W} \frac{\aCz}{\Lambda^2} F_{\mu\alpha} F^{\mu\beta} Z^\alpha Z_\beta,\nonumber 
      \end{eqnarray}
where in addition to the convention already introduced, $\Lambda$ denotes the energy scale where a new physics is assumed to appear and $\Theta_W$ is the
 Weinberg angle. Note that the general parametrization of the $\wwgamma$ and $\gamma\gamma WW$
Lagrangians can be found in~\cite{Belanger:1992qh,Hagiwara:1986vm}.

\end{document}